\documentclass[aas_macros]{elsart4-1}

 \usepackage{graphicx}

\usepackage{amssymb}
\usepackage{amsmath}
\usepackage{wrapfig}
\usepackage{bbm}
\usepackage[english,francais]{babel}
\usepackage[latin1]{inputenc}

\newtheorem{e-proposition}[theorem]{Proposition}

\newtheorem{e-definition}[theorem]{Definition\rm}

\renewcommand{\theequation}{\arabic{equation}}
\def\og{\leavevmode\raise.3ex\hbox{$\scriptscriptstyle\langle\!\langle$~}}
\def\fg{\leavevmode\raise.3ex\hbox{~$\!\scriptscriptstyle\,\rangle\!\rangle$}}

\newcommand{\be}{\begin{equation}}
\newcommand{\ee}{\end{equation}}

\newcommand{\dg}{$^\circ$}

\newcommand{\pcc}{\textrm{cm}\ensuremath{^{-3}}}

\newcommand{\muG}{\ensuremath{\mu}\textrm{G} }
\newcommand{\degree}{\ensuremath{^\circ}}

\newcommand{\ncre}{$n_{\rm cre}$}
\newcommand{\neth}{$n_{e}$}

\begin{document}

\begin{frontmatter}


\title{The Galactic Magnetic Field and \\ Ultrahigh-Energy Cosmic Ray Deflections}
\author{Glennys R. Farrar}
\ead{gf25@nyu.edu}
\address{Center for Cosmology and Particle Physics, Department of Physics, New York University, USA}
\selectlanguage{english}

\medskip
\begin{center}
\end{center}

\begin{abstract}
Our understanding of the Galactic magnetic field (GMF) has increased considerably in recent years, while at the same time remaining far from adequate.  By way of illustration, the Jansson and Farrar (2012) (JF12) GMF model is described, emphasizing how it is constrained and which features are robust or likely to change, as modelling and constraining data improve.   The most urgent requirements for the next phase of modelling are  a more realistic model for the relativistic electron distribution (in order to reduce the systematic error associated with interpreting synchrotron data) and a better theoretical understanding of the origin of the large-scale coherent field (in order to develop a better phenomenological parameterization of the field).  Even in its current stage of development, the JF12 model allows some important conclusions about UHECR deflections in the GMF to be formulated.   
{\it To cite this article: Glennys R. Farrar, C. R. Physique XX (2014).}

\vskip 0.5\baselineskip

\selectlanguage{francais}
\noindent{\bf R\'esum\'e}
\vskip 0.5\baselineskip
\noindent
{\bf Le champs magnétique Galactique et la déflexion des rayons cosmiques ultra-énergétiques. }
Notre compréhension du champs magnétique Galactic (GMF) s'est considérablement améliorée au cours des dernières années mais reste largement insuffisante.  A titre d'illustration le modèle GMF de Jansson et Farrar (2012) (JF12) est décrit en insistant sur la manière dont il est contraint et sur ses caractéristiques, qu'elles soient robustes ou, au contraire, susceptibles de changer avec l'amélioration de la modélisation et des données. Les besoins les plus urgents pour la prochaine phase de modélisation sont, d'une part, un modèle plus réaliste de la distribution des électrons relativistes (ce qui permettra de réduire les incertitudes systématiques associées à l'interprétation des données d'émissions synchrotron)  et, d'autre part, une meilleure compréhension théorique de l'origine du champ Galactique cohérent sur les grandes échelles (afin de développer une meilleure paramétrisation phénoménologique du champs). Le modèle JF12, même dans sa version actuelle, permet de formuler quelques conclusions importantes sur la déflexion des RCUHE dans la Galaxie. 
{\it Pour citer cet article~: Glennys R. Farrar, C. R. Physique XX (2014).}

\keyword{UHECR; Galactic Magnetic Field; magnetic deflections} \vskip 0.5\baselineskip
\noindent{\small{\it RCUHE~:} Champs magnétique Galactique~; déflexions magnétiques}}
\end{abstract}
\end{frontmatter}

\selectlanguage{francais}
\section*{Version fran\c{c}aise abr\'eg\'ee}

\selectlanguage{english}
\section{Introduction}
\label{intro}

Our understanding of magnetic fields in the Milky Way and of the global structure of the GMF has developed over many decades.  A recent review of the magnetic fields in galaxies with a discussion of observations is that of Beck and Wielebinski (2013) \cite{beckWieleb13}; a classic review of the previous generation of GMF models and their observational constraints, and application of dynamo theory to interpret the observations is Beck, Brandenburg, Moss, Shukurov and Sokoloff (1996) \cite{beckAnnRevAst96}.   It has only been in the past few years that the present generation of more sophisticated and quantitatively-constrained models of the GMF have emerged.  

\begin{figure}[b]
\centering
\vspace{+0.05in}
\includegraphics[width= \textwidth]{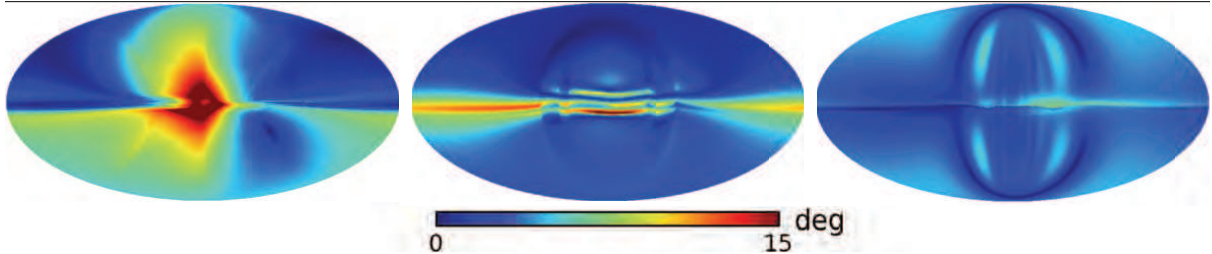}
\caption{The magnitude of the deflection of a 60 EeV proton, displayed by arrival direction, in (left to right) the JF12, SR10 and PTK11 models of the large-scale GMF.}\label{magdefscompare}
\vspace{+0.1in}
\end{figure}

Even with these improved models, predicting UHECR deflections remains uncertain, as seen from Fig. \ref{magdefscompare} which shows the magnitudes of UHECR deflections predicted by three recent GMF models due to Sun and Reich (SR10, based on \cite{sun+08} incorporating the parameter update of \cite{sr10}), Pshirkov, Tinyakov and Kronberg (PTK11,\cite{pshirkov+11}) and Jansson and Farrar (JF12,\cite{jf12a}).  Evidently, determining the astrophysical source(s) of UHECRs via their correlations with candidate source catalogs is fraught with uncertainty, even if the charge and energy of each UHECR were perfectly known, until we know the GMF with greater confidence.  Nonetheless some general expectations for UHECR deflections are emerging, as discussed below.

Trustworthy 3D models of the GMF are also needed to calculate the propagation of Galactic cosmic rays and predict Galactic diffuse gamma emission, with wide-ranging implications including interpreting possible signatures of Dark Matter annihilation in the Galaxy.  Another important application is modeling the spatial dependence of the synchrotron emission spectral index, needed for accurate foreground subtraction to obtain the cosmological CMB signal.

\section{Observables to constrain the GMF}
\label{observables}
Faraday Rotation Measures (RMs) of extragalactic sources have always been primary constraining data for Galactic magnetic fields.  Rotation Measures depend on the line-of-sight field;  in units of rad $\textrm{m}\ensuremath{^{-2}}$, 
\begin{equation}
\textrm{RM} = 0.81\int_0^{L}\left(\frac{n_e(l)}{\pcc} \right)
\left(\frac{B_\parallel(l)}{\muG} \right) \left(\frac{{\rm d} l}{\textrm{pc}}
\right),
\label{eq:RM}
\end{equation}
where $L$ is the distance to the source and $n_e$ is the density of ionized electrons, which is dominated by the \emph{thermal} electron density.  The sheer number of extragalactic sources providing all-sky, well-measured RMs is roughly 2 orders of magnitude larger than in the early 1990s, and will increase by another 1 or 2 orders-of-magnitude within the decade.  

Each extragalactic source (primarily distant quasars) has an intrinsic RM due to magnetic fields in the AGN and its host galaxy, which can be large compared to the Galactic contribution.  However the extragalactic contributions are uncorrelated from one source to another, so their impact is to produce an approximately isotropic variance beyond that due to random fields in the Galaxy.   Increasing the number of extragalactic sources in the RM dataset benefits the GMF constraints by reducing the fluctuations in the mean RM of each analysis pixel. 

Complementing RM, we now have all-sky maps of polarized and unpolarized Galactic synchrotron emission from WMAP and soon also from Planck, as well as targeted observations of Galactic structures and high resolution synchrotron mapping of external galaxies such as in the CHANG-ES survey\cite{CHANG-ES}.   Galactic synchrotron emission provides a constraint on the GMF which is complementary to that from RMs, because synchrotron emission depends on the transverse magnetic field, weighted by the relativistic (also called cosmic ray) electron density, $n_{\rm cre}$.  The polarization state of linearly polarized light is specified by the Stokes parameters Q and U, with each proportional to the polarized intensity (PI) and the relation between Q and U encoding the orientation of the transverse component of the coherent magnetic field (c.f., \cite{rybickiLightman}), with
\begin{equation}
\textrm{PI} \sim \int_0^{L} \, n_{\rm cre}(l) ~
B^{2}_\perp (l) \, {\rm d} l
.
\label{eq:QU}
\end{equation}

Exploiting the relationships in Eqs. (\ref{eq:RM}) and (\ref{eq:QU}) between the magnetic field and the physical observables depends on knowing the thermal and relativistic electron distributions.  The standard model for \neth\ is the Cordes-Lazio NE2001 model \cite{NE2001}, with the mid-plane density and vertical scale-height modified according to \cite{Gaensler:2008}.  The primary observational constraint on NE2001 comes from the dispersion measures (DMs) of pulsars, proportional to the column density of free electrons along the line of sight to the pulsar; the DM is obtained from the frequency dependence of the arrival time delay of a broadband pulse.    The relativistic electron distribution is much more uncertain.  JF12 considered the two available models and found that the GALPROP distribution (kindly provided by A. Strong), with an overall rescaling parameter $\alpha$, gave the best fit;  the fit could be improved by more extensive modifications, e.g., by changing the vertical scale height, but exploring that domain was left for the future.

Combining RMs and DMs of Galactic pulsars in principle can provide a very high-value constraint\cite{han+pulsars06}. Firstly, pulsars have negligible intrinsic RM, and even more importantly, since pulsars are {\em inside} the Galaxy, if their distance is known that adds longitudinal resolution in the constraints from Eq. (\ref{eq:RM}).  However until recently, only a handful of pulsars have had a well-measured distance, as is necessary for constraining the large scale field given the field reversals in the Galactic disk; fortunately, the number of pulsars with parallax distances and RMs is growing rapidly now.  (Note, however, that correlations between turbulence-induced fluctuations in $n_{e}$ and $B$ can result in biases\cite{beckAA03}.)   Dramatic improvements are also unfolding in RM synthesis modeling, such as in the GMIMS survey \cite{Wolleben:2010}, which will lead to better understanding of foreground random fields. 

\section{The Galactic magnetic field}
\label{JF12}  

Thanks to nearly all-sky RM and synchrotron emission data, and modern computational resources, some general features of the GMF have emerged.  We discuss below the JF12 field model \cite{jf12a,jf12b}, describing how it was constrained and noting its deficiencies.  This example should provide guidance to the next stage of theoretical understanding and phenomenological modeling.  

\subsection{Jansson-Farrar JF12 model}
The Jansson-Farrar JF12 model \cite{jf12a,jf12b} has three distinct parts: i) the coherent large-scale field, with disk, halo and out-of-plane components\cite{jf12a},  ii) a fully random field specified by its spatially-varying rms field strength \cite{jf12b}, and iii) a ``striated'' random field, defined and explained below\cite{jf12a}.  The notions of coherent and random fields have no {\em a priori} boundary, but practically speaking the coherent field is the component of the field which can be described as a relatively simple function of position for the entire Galaxy, while the random field averages to zero on scales larger than a few hundred pc or less.  The original JF12 analysis\cite{jf12b} constrains only the RMS value of the random field and does not determine its maximum coherence length or power-spectrum behavior.\\

\begin{figure}
\centering
\includegraphics[width=0.43 \textwidth]{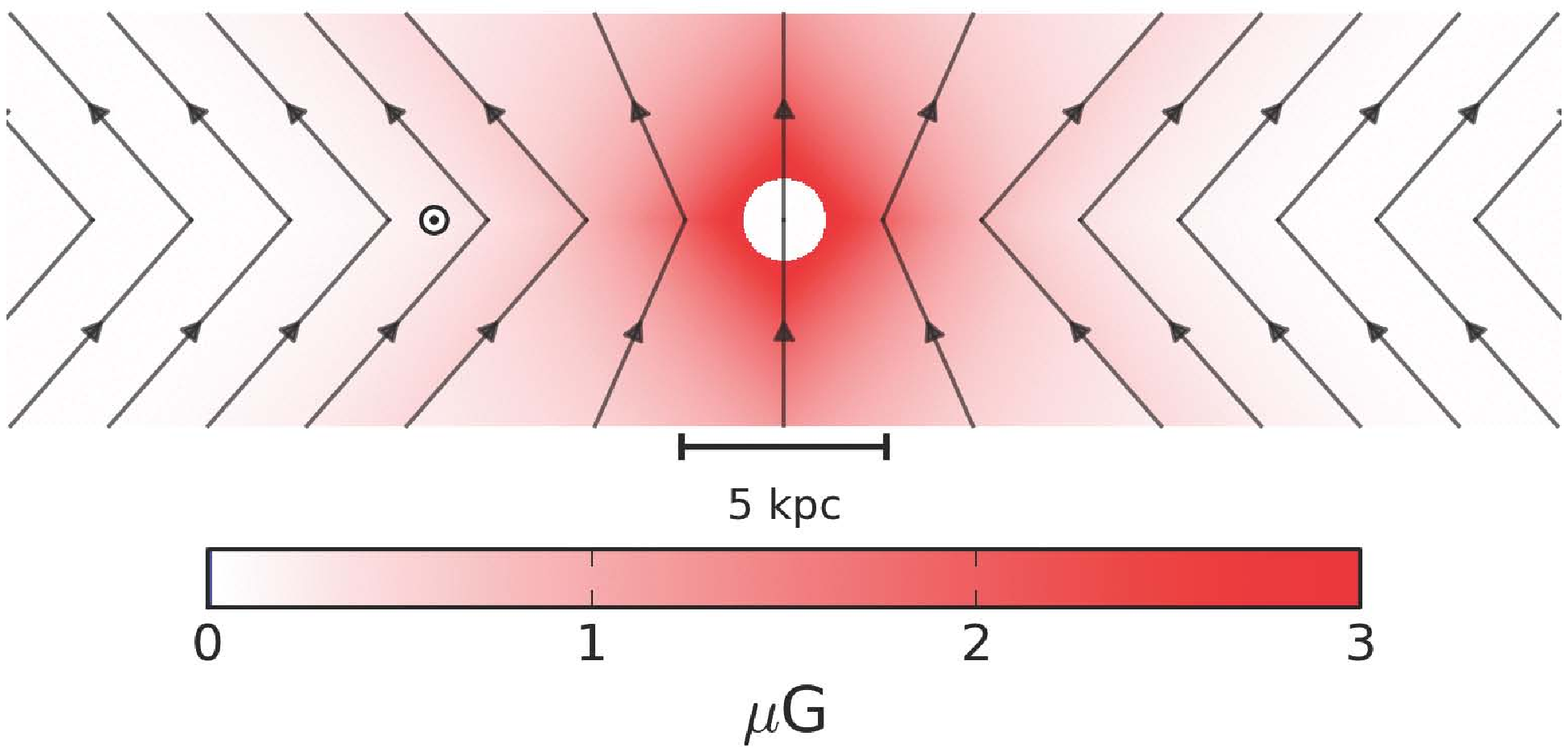}
\quad
\includegraphics[width=0.45 \textwidth]{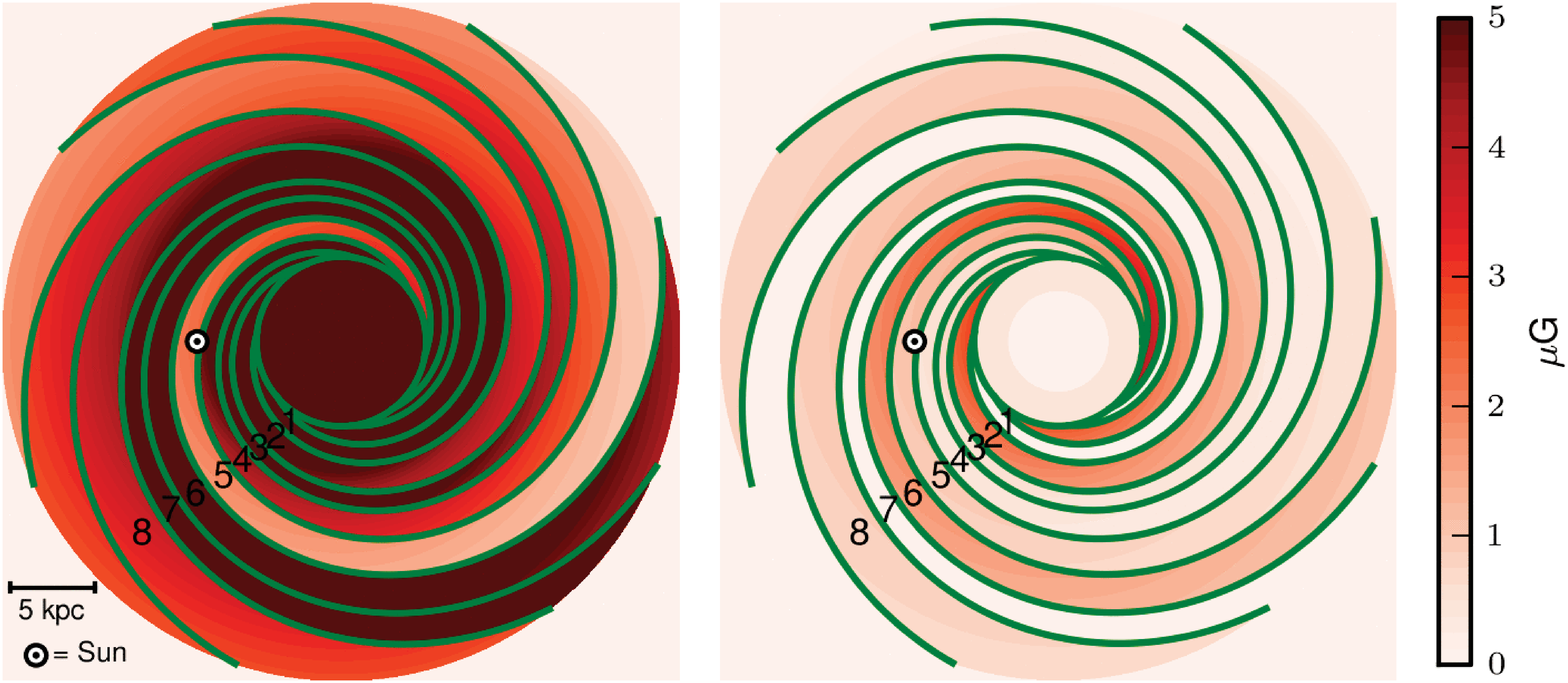}
\caption{Left:  An $x-z$ slice of the galaxy showing the geometry of the out-of-plane ``X'' component;  the oriented lines show the direction of the field and do not depict the field strength which is indicated by the shading. Right: The disk components of the JF12 rms random field \cite{jf12b} and JF12 regular field (clockwise in rings 3-6 and counterclockwise in 1,2,7, and 8)\cite{jf12a}, respectively; the same color code is used for both to facilitate comparison.}\label{JF12B}
\vspace{-0.01in}
\end{figure}

\noindent \underline{Coherent field}\\
The disk component of the JF12 coherent field was taken to be toroidal in an inner ``molecular ring'' region from 3-5 kpc, beyond which it has a logarithmic-spiral geometry adopted from \cite{Brown:2007}.  Thus the coherent disk field has 10 parameters -- the value in the (toroidal) molecular ring region, 7 field values at 5 kpc (the 8th being fixed by flux conservation), and 2 parameters specifying the thickness profile of the disk.  The random and coherent fields in the disk found by JF12 are shown in the right panel of Fig. \ref{JF12B}. 

The toroidal halo component of the coherent field is described by 6 parameters:  the overall strength (including direction) and radial fall-off, independently in the northern and southern hemispheres, and the vertical scale height and radial transition width, taken to be the same in N and S to limit the number of free parameters.  Alternate forms were explored including a vertical extension of the disk's spiral-arm structure in which the pitch angle and phase of the spiral was allowed to change with ${z}$, but a purely toroidal field was found to give the best fit. 

Finally, an out-of-plane halo component of the coherent field designated as the ``X-field'' was included, partially motivated by the X-shaped  field structures seen in  radio observations of external, edge-on galaxies \cite{Krause:2009,Beck:2009}.  This component was taken to be azimuthally symmetric and poloidal, i.e., the X field has no azimuthal component, that being incorporated via the toroidal halo field. The disk, toroidal halo and X fields were required to be separately divergenceless so their parameters could be changed independently.   The parameterization adopted for the X field in JF12 has 4 free parameters: the value at the origin (where by azimuthal symmetry it is purely in the $\pm \hat{z}$ direction), the radial scale-length of the field strength, the rate at which the field lines open in the central region, and the angle they make to the disk at asymptotically large radius or equivalently the radius beyond which the angle stops changing.  A visualization is provided in Fig. \ref{JF12B} for the parameters of the best-fit GMF. \\

\noindent \underline{ Fully random field}\\
The random field is taken to be a superposition of a disk component whose spiral-arms are the same as used for the coherent field, but with independently fitted rms field strength, and a smooth halo component.  The latter does not have any azimuthal structure; it is characterized by its overall rms strength, and radial and vertical scale lengths.  The disk component has 11 free parameters -- one more than for the coherent field where flux conservation provides one constraint.\\

\noindent  \underline{ Striated field}\\
The striated component is a new element in the JF12 analysis.  Phenomenologically, it allows for the possibility of a random field which on large scales averages to zero but which is locally aligned in some preferred direction. Such a ``striated'' field would arise from stretching or compressing a fully random field, which was the motivation for including it.  JF12  found that, within the discriminating power of their analysis, the striated field is aligned with the local coherent field and its rms strength is locally proportional to the coherent field strength: $B^2_{\rm stri} =\beta B_{\rm reg}^2$.  Several variations to this simple model were tried and none gave as good a fit, but an exhaustive exploration of the striated random component of the field was not undertaken.  An alternate physical origin for such a field component is the local {\em compression and rarification} of a coherent field, as occurs when a supernova explosion produces a shell of compressed material surrounding a rarefied plasma.  This has the phenomenological impact of increasing polarized synchrotron emission relative to that expected in a smooth medium without impacting the RM prediction (in the approximation that the thermal electron density is slowly varying), since synchrotron emission is quadratic in the field strength and RM is linear.  Future more detailed studies of the relation between striated and coherent fields as a function of position in the Galaxy will help ascertain the origin of this contribution to the Galactic synchrotron signal.\\

\begin{figure*}
\vspace{-0.25in}
\centering
\includegraphics[width=0.95 \linewidth]{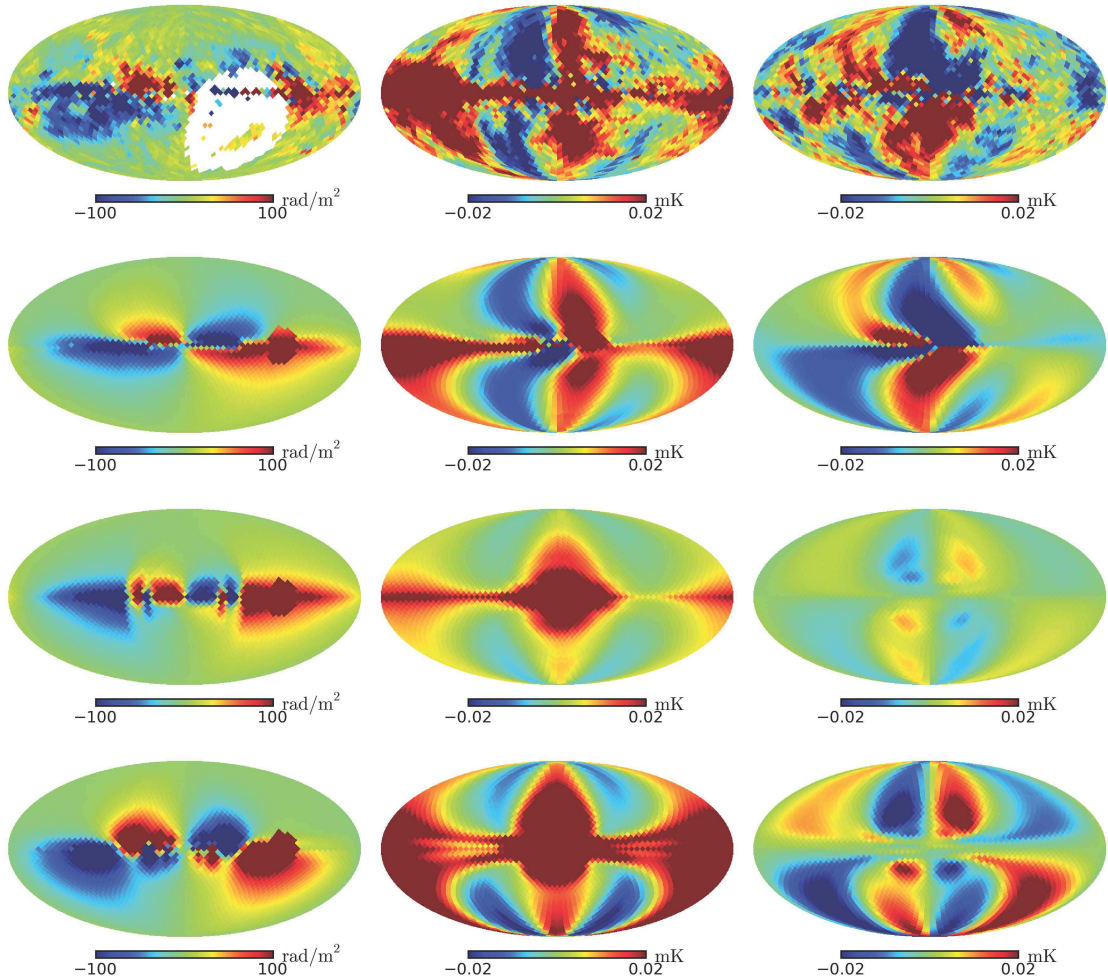}
\vspace{-0.05in}
\caption{Skymaps in Mollweide projection of (left to right) RM (in rad/$m^2$), Stokes $Q$ and Stokes $U$ (in mK).  From top to bottom: row 1, data; rows 2-4 simulated data from the JF12, SR10 and PTK11 models, respectively.  Galactic longitude $l=0\degree$ in the center and increases to the left.  White pixels (RM) lack data.  The $Q,\,U$ data were not fit in a masked region where foreground contamination is greatest, yet the JF12 fit is quite good in all regions; see \cite{jf12a} for the masks used.  Plot courtesy D. Khurana.
} \label{data}
\end{figure*}

\noindent \underline{ Fitting procedure and results}\\
The full JF12 model thus depends on 35 parameters: 20 for the coherent field, one for the striated field denoted $\beta$, 13 for the random field, and the overall rescaling of $n_{\rm cre}$, denoted $\alpha$.  These parameters were constrained by Markov Chain Monte Carlo (MCMC) fitting altogether nearly 10,000 observables, each with measured variance, distilled from 40,000 extragalactic RMs, and the WMAP polarized synchrotron and total synchrotron emission maps, Q, U, and I.  Only the coherent field contributes to the RMs, while both regular and striated fields contribute to Q and U.  First the parameters of the coherent field and the product $\alpha  \beta$ are fit to the RM, Q and U data, then the total synchrotron emission I is fit, fixing the parameters of the random field and breaking the degeneracy between $\alpha$ and $\beta$.  The best-fit parameters can be found in refs. \cite{jf12a,jf12b} where it can be seen that with the assumed GMF functional form and $n_{e}$ and $n_{\rm cre}$ models adopted, all parameters are well-constrained except for the radial transition width of the southern toroidal halo. 

Fig. \ref{data} shows the RM, Q and U data compared to the maps predicted by the JF12, SR10 and PTK11 models;  corresponding plots for the JF12 random field are given in \cite{jf12b}.   The greater fidelity seen with the JF12 model is related to its functional form which is divergenceless and allows for possible out-of-plane and striated components (see \cite{jf12b} for a comparison of the $\chi^{2}$ per degree of freedom for the different model fits shown in Fig. \ref{data}).  The X-field specifically is responsible for producing the L-R asymmetry and sharp boundaries seen in the Q \& U maps.  The striated component of JF12 effectively enhances the synchrotron emission, for a given field strength and \ncre.  

\subsection{Uncertainties and caveats in GMF models}

\begin{figure}
\centering
\vspace{-0.3in}
\begin{minipage}[b]{0.49 \textwidth}
\vspace{-0.5in}
\includegraphics[width= \textwidth]{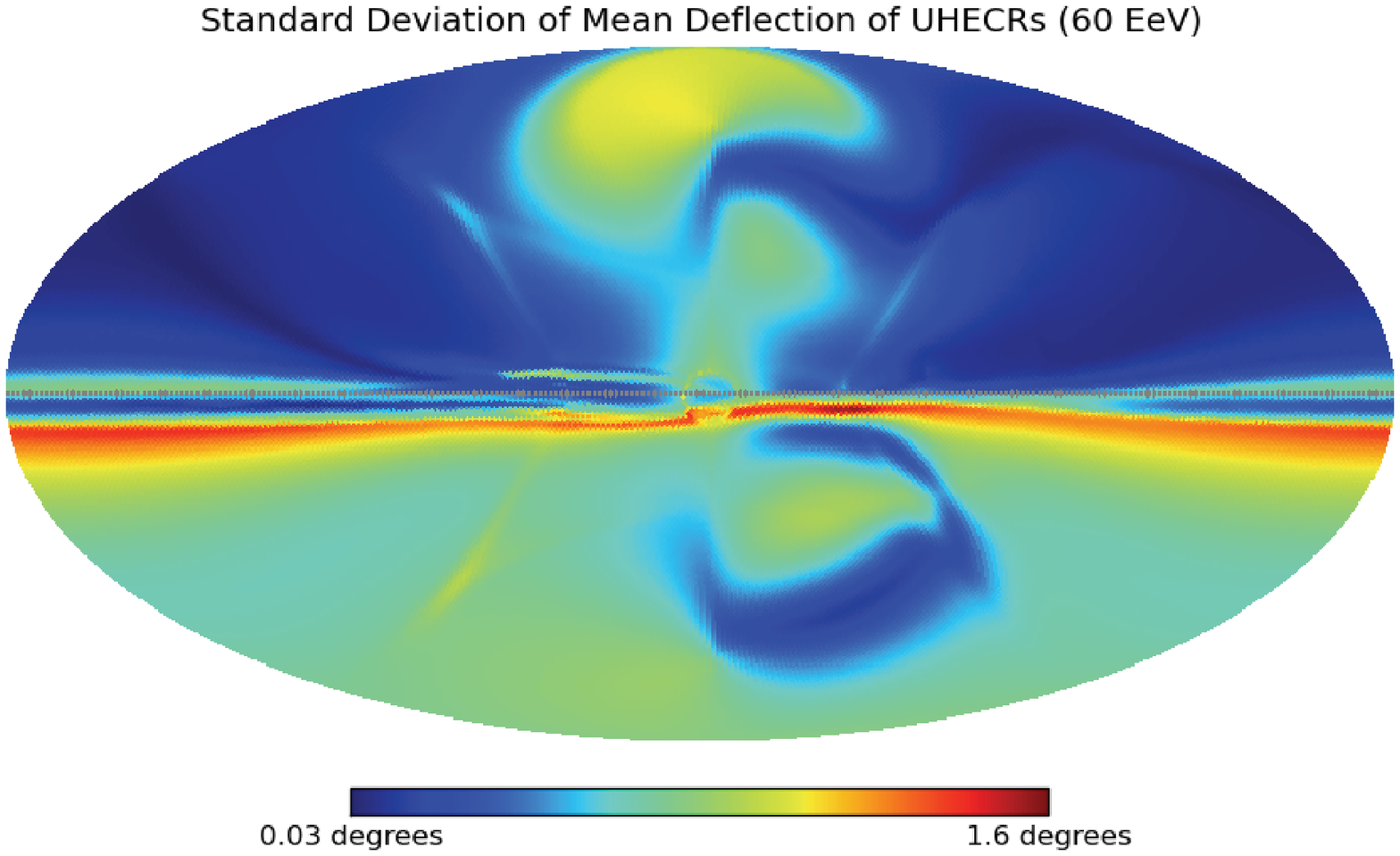}
\vspace{-1.5in}
\caption{The {\em uncertainty} in degrees on the deflections of 60 EV UHECRs shown in Fig. 1a, stemming from the parameter uncertainties in the coherent JF12 GMF, plot courtesy D. Wilmes. }\label{wilmes}
\end{minipage}
\quad
\begin{minipage}[b]{0.48 \textwidth}
\vspace{-0.4in}
\includegraphics[width= \textwidth]{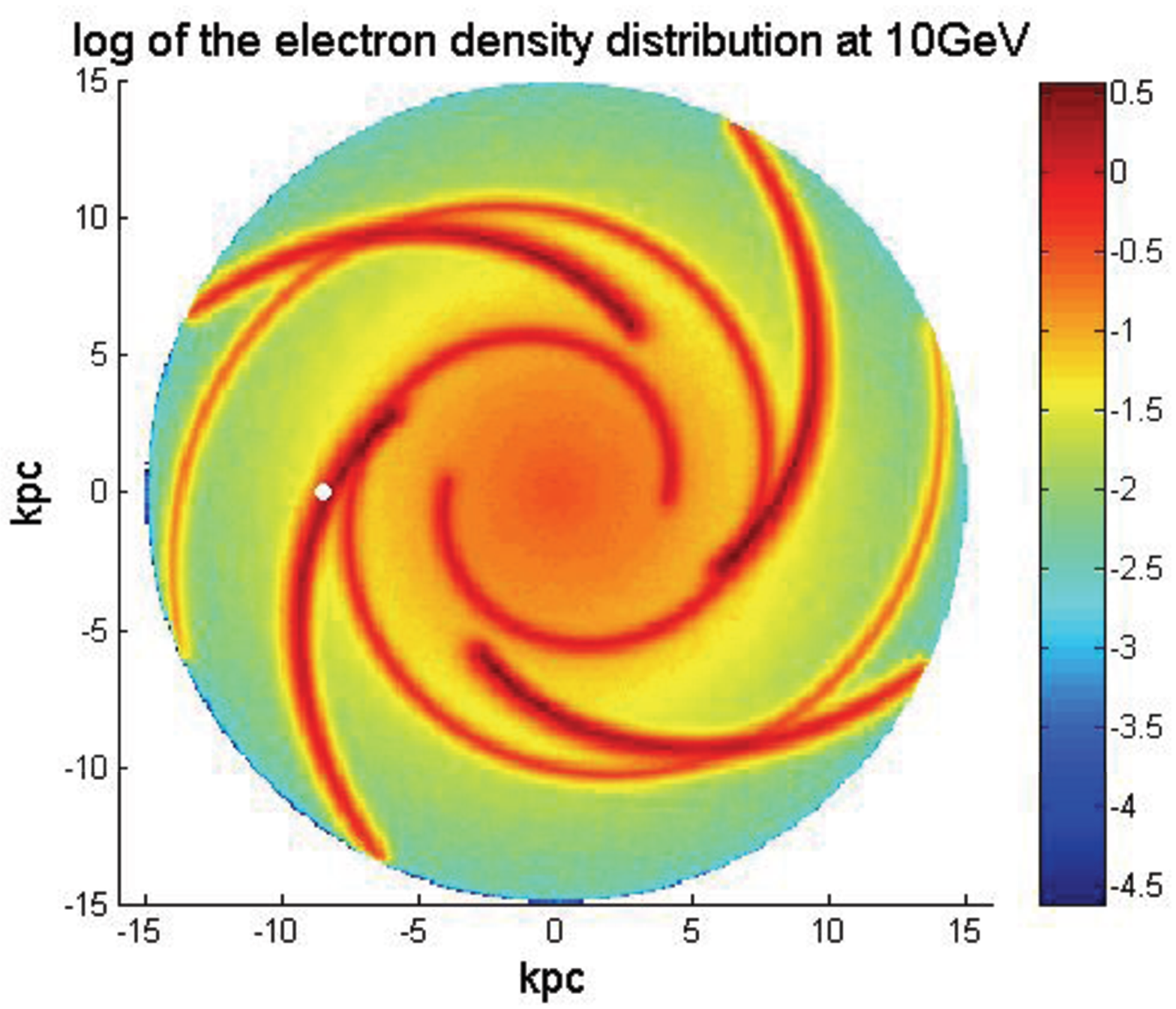}
\vspace{-0.3in}
\caption{The distribution of 10 GeV electrons, ${ln(n_{\rm cre}/n_{\rm cre,\odot})}$, in the model of ref. \cite{benyamin+13}, courtesy D. Benyamin.} 
\label{ncreBenyamin}
\end{minipage}
\vspace{-0.05 in}
\end{figure}

The MCMC procedure used in JF12 to determine the parameters of the GMF returns a multi-dimensional probability distribution for the parameter values, which reflects correlations in the uncertainties that are hard to appreciate looking at individual parameter uncertainties alone.  To have a practical measure of how well constrained a GMF model is, we calculate the uncertainty in UHECR deflections due to uncertainties in the model parameters, by randomly sampling the multi-dimensional parameter distribution returned by the MCMC analysis.  Fig. \ref{wilmes} shows a map of the variance in the deflections thus obtained, as a function of the CR arrival direction, for the JF12 field model at a rigidity of 60 EV.  This shows that the UHECR deflection-mapping uncertainties stemming from the analysis and fitting procedures and from the variance in the underlying constraining data are quite low, with a maximum value of $1.6^{\circ}$.  Thus by reducing the uncertainty in the \ncre\ and $n_{e}$ models and improving the functional form of the GMF model, in principle quite reliable deflection mapping can be eventually achieved, at least for the rigidities of protons subject to the Greisen Zatsepin Kuzmin (GZK) energy loss horizon.  In any case, the present uncertainty in the GMF is much larger than that stemming from the fitting, exemplified in Fig. \ref{wilmes}.  We now turn to the caveats that must be placed on the validity of the JF12 model, and more generally to any effort to constrain the large scale structure of the GMF.  

As stressed in \cite{jf12a} and above, having a correct \ncre\ and $n_{e}$ model is essential for obtaining a correct $\vec{B}$ field, yet the azimuthally symmetric GALPROP \ncre\ model may be subject to major revision, as illustrated in Fig. \ref{ncreBenyamin} showing the (natural log of the) 10 GeV electron density in the Galactic plane relative to the solar neighborhood, in the model of \cite{benyamin+13}.  This is the characteristic energy of electrons producing the 22 GHz synchrotron emission mapped by WMAP and used in the JF12 constraints.  In the model \cite{benyamin+13}, sources are predominantly pulsars in the spiral arms, and, with expected diffusion parameters, electrons cool before traveling far enough to smooth out the spiral arm structure; Fig. \ref{ncreBenyamin} exhibits azimuthal modulations of nearly two orders of magnitude.  Furthermore,  the spiral arm geometry used in \cite{benyamin+13} is significantly different from the arm geometry of the standard Cordes-Lasio \neth\ model, NE2001, used by \cite{Brown:2007} and adopted by JF12.  

The possible impact on the JF12 random field model of converting to the \ncre\ model of Fig. \ref{ncreBenyamin}, can be very roughly estimated by idealizing the change to \ncre\ as being to replace the azimuthally uniform GALPROP value, $n_{G}$, by 0 in the interarm regions and by the value $n'$ in the arms.  Let $f$ be the fraction of the line of sight occupied by arms.  Reproducing the observed synchrotron emission would then require a reduction in the value of $B_{rms}$ by a factor $\approx \sqrt{ (n_{G}/ n' )\,f^{-1}}$.  From Fig. \ref{ncreBenyamin}, $f \approx 0.1$ and $n'/n_{G} \approx 10^{2}$, leading to the rough estimate that the peak rms fields in the disk, in the JF12 random field model, may need to be reduced by a factor of order 3.  

The above crude estimate may be indicative of the impact on the disk random field, of converting from the smooth GALPROP \ncre\ model to a structured one such as given by \cite{benyamin+13}, but the impact on the coherent and striated fields in the disk cannot be estimated in this way since those are constrained in part by RMs which are not sensitive to \ncre.   Some modification should also be expected to the halo and $X$ fields, since all lines-of-sight receive a contribution from the disk, albeit a small one for higher galactic latitudes which provide the strongest constraints on those components.  

Clearly, a high priority for next-generation GMF modeling, is to fit for the arm geometry as well as field parameters in the disk.  The arm geometry of \cite{benyamin+13} is suggested by spiral density wave theory\cite{LinShu64} with parameters chosen based on CO (carbon monoxide) maps.   Simultaneously fitting for parameters of the \ncre\ and $n_{e}(\vec{r})$ models, including pertinent additional data such as DMs, may be feasible and is being explored. 

\subsection{Summarizing the status of the GMF}

\noindent To recapitulate, the key features of the JF12 halo field are: \\
$\bullet$ A North-directed, coherent poloidal component (the X-field).  Its strength is $\approx 5 \mu$G at the center of the galaxy and diminishes relatively slowly with distance from the Galactic plane;  its radial scale-length is $\approx 3$kpc and its value in the solar neighborhood is $\approx 0.2 \mu$G.  \\
$\bullet$ Oppositely directed coherent toroidal fields above and below the Galactic Plane.  The sense of rotation below the plane (Southern hemisphere) is in the same direction as the rotation of the disk.  The toroidal fields reach their maximum strength about 1 kpc away from the plane, beyond which they decline slowly reaching half their peak value at $\approx 5$ kpc.  The sense of the toroidal fields are consistent with their resulting from differential rotation of the coherent poloidal X-field.\\
$\bullet$ The halo component of the random field declines slowly with radius and $|z|$, with a value $1 \mu$G for $|z| \approx 3$ kpc at the solar circle.

In spite of the caveats about sensitivity to the uncertain \ncre\ distribution, the following gross features of the JF12 disk field are probably robust:  
\\
$\bullet$ There are two main arms with oppositely-oriented fields and the solar system is in an inwardly-pointing arm.  Note that at the solar radius, the assumed logarithmic spiral geometry is almost azimuthally oriented.  
\\
$\bullet$ The typical strength of the coherent disk field in the magnetic arms is  $\approx 1 \mu$G, with maximum values of a few $\mu$G.  In the disk, the rms strength of the random field is considerably larger than of the coherent field, with a peak strength of order 10$\mu$G, with large uncertainty due to the uncertainty in $n_{e}(\vec{r})$.

The large scale poloidal and toroidal halo fields of the Galaxy provide a valuable hint for understanding their origin, but this topic is beyond the scope of this Chapter.  

\section{Deflections of UHECRs in the Galaxy}

\begin{wrapfigure}{l}{0.38\textwidth}
\vspace{-0.15in}
\centering
\includegraphics[width=0.4\textwidth]{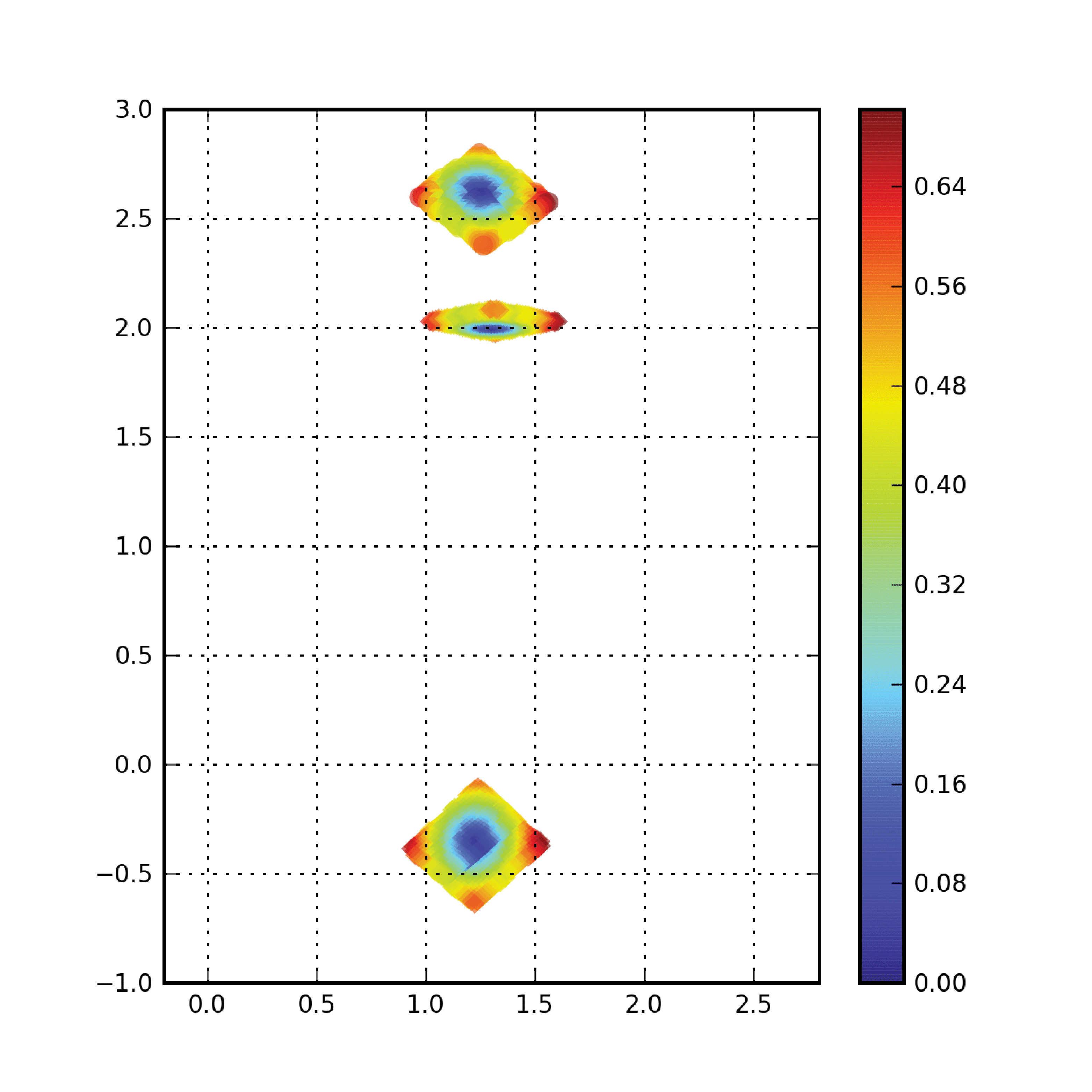}
\vspace{-0.25in}
\caption{The arrival plane regions contributing to detected 60 EV CRs for a source at \{l=341.7\dg, b=4.2\dg\}.  Photons arrive at (0,0) and distances are in kpc; color shows the angular distance of the UHECR from the center of its image.}\label{arrivalplane}
\end{wrapfigure}
\noindent Although JF12 cannot be trusted in detail, it should be good enough to draw a number of qualitative conclusions about deflections of the UHECRs which do not pass too close to the disk.   A very interesting feature 
to be expected of UHECR propagation in the GMF was pointed out  and investigated by Harari, Mollerach 
and Roulet in a series of papers beginning in the 1990s, e.g., 
\cite{hmrLensing00}.  Namely, the GMF acts as a lens whose properties change with rigidity\footnote{Rigidity, energy divided by charge, is all that matters for cosmic ray deflections.  Rigidity is properly measured in units of V but when only magnetic deflections and not energy losses are of concern, V and eV maybe used interchangeably since knowing the deflection of a proton of energy E in eV specifies the deflection of any CR with the same value of E/Z.}, in principle giving rise to multiple images and magnification.  
We can take the JF12 model as a semi-realistic magnetic field model to explore whether the rigidities at which multiple imaging and magnification occur are high enough for these phenomena to be observable.  Another important question is whether the random field seriously modifies deflections compared to those in the regular field alone.  As we shall see, the consequences of GMF deflection can be dramatic and will need to be taken into account in any careful future effort to do source-identification.  In the following discussion, extragalatic deflections will be ignored, but in a comprehensive study their impact also needs to be considered.

While it is easy to accept the idea of multiple-images and image distortions based on everyday experience, sometimes people get confused by a seeming paradox which may be encountered in thinking about UHECR deflections, so we will begin with discussing the origin of the imaging.  UHECRs from a particular distant source arrive to the Galaxy as a uniform, parallel beam.  Define an ``arrival plane'' transverse to the direction to the source, where the GMF starts to be relevant.  The subsequent trajectory of a UHECR in the GMF will depend on its rigidity and where it intersects the arrival plane.   Most locations on the arrival plane do not illuminate the Earth at all -- those UHECRs' trajectories simply miss Earth altogether.  It can happen that, for a sufficiently low rigidity or in a complex enough magnetic field, more than one region of the transverse plane illuminates the Earth.  Fig. \ref{arrivalplane} shows this phenomenon by showing the position on the arrival plane, of CRs which hit the detector from a given source; their diamond-shapes reflect the pixelazation of the arrival direction binning.

\begin{figure}[t]
\hspace{-0.5in}
\centering
\begin{minipage}[b]{0.48 \textwidth}
\vspace{-0.2in}
\includegraphics[width= \textwidth]{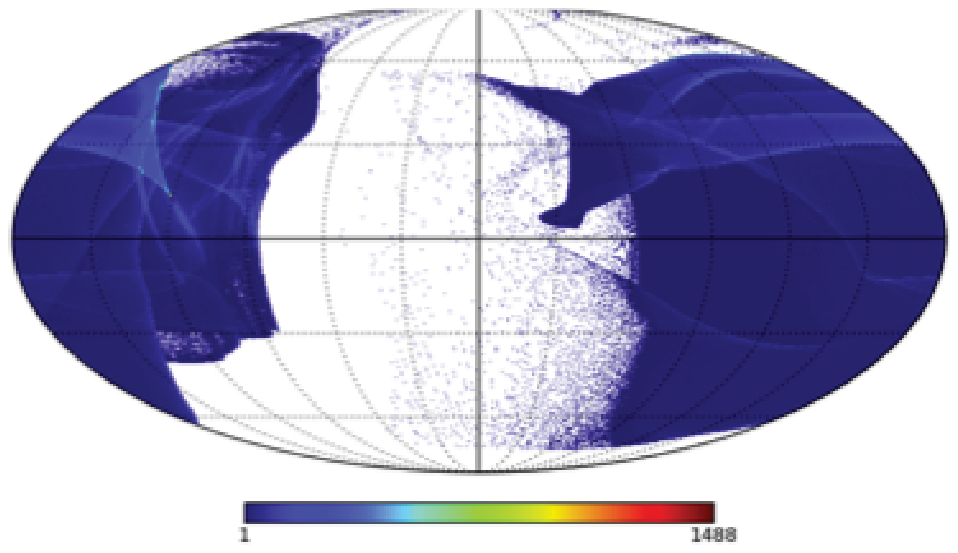}
\end{minipage}
\begin{minipage}[b]{0.48 \textwidth}
\vspace{-0.3in}
\includegraphics[width= \textwidth]{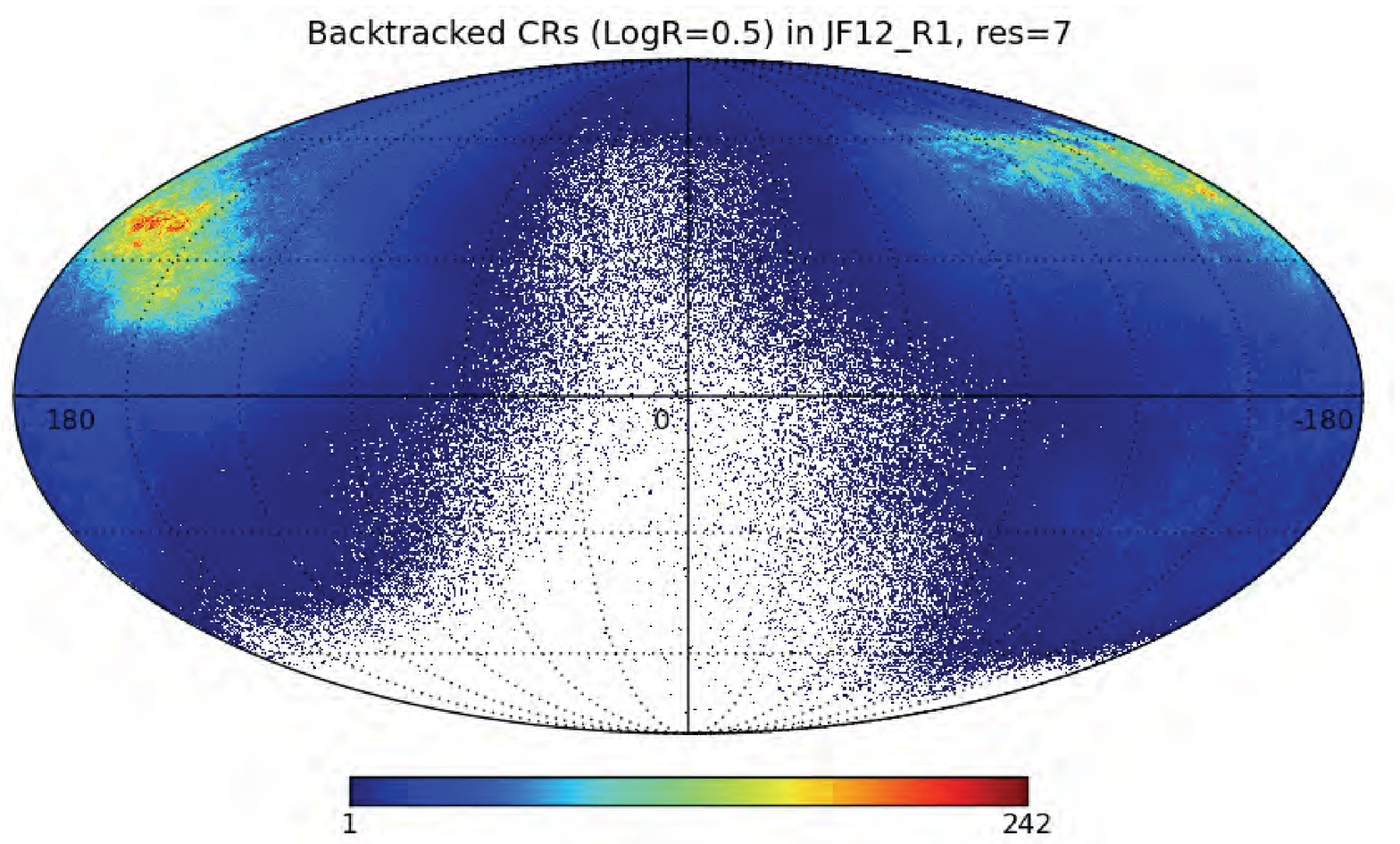}
\end{minipage}
\vspace{-0.15in}
\caption{Magnification map at 3.2 EV for JF12 regular (left) and regular+random fields (right).  } 
\vspace{-0.05in}
\label{magmaps}
\end{figure} 

It is intuitively plausible, and required by the conservation of specific intensity along a CR trajectory (flux per steradian, c.f., \cite{rybickiLightman}), that CRs in each of the non-overlapping patches follow distinct trajectories, such that in the limit of infinite angular resolution they arrive in non-overlapping images.  As the rigidity decreases, the complexity of the trajectories and the number of images increases, although within the detector resolution, images may merge.   It should be noted that the size of the various regions in the arrival plane which illuminate the detector are in general not equal.   Since the flux is uniform in the arrival plane, the relative numbers of events in the various detected images are proportional to the area in the arrival plane which produces them, or equivalently, to the angular size of the given arrival-direction image.  This means that the total flux from a given source can be magnified or demagnified relative to the total flux that would be received from the source in the absence of the GMF.

\begin{figure}[b]
\hspace{-0.3in}
\centering
\begin{minipage}[b]{0.48 \textwidth}
\includegraphics[width=1.15 \textwidth]{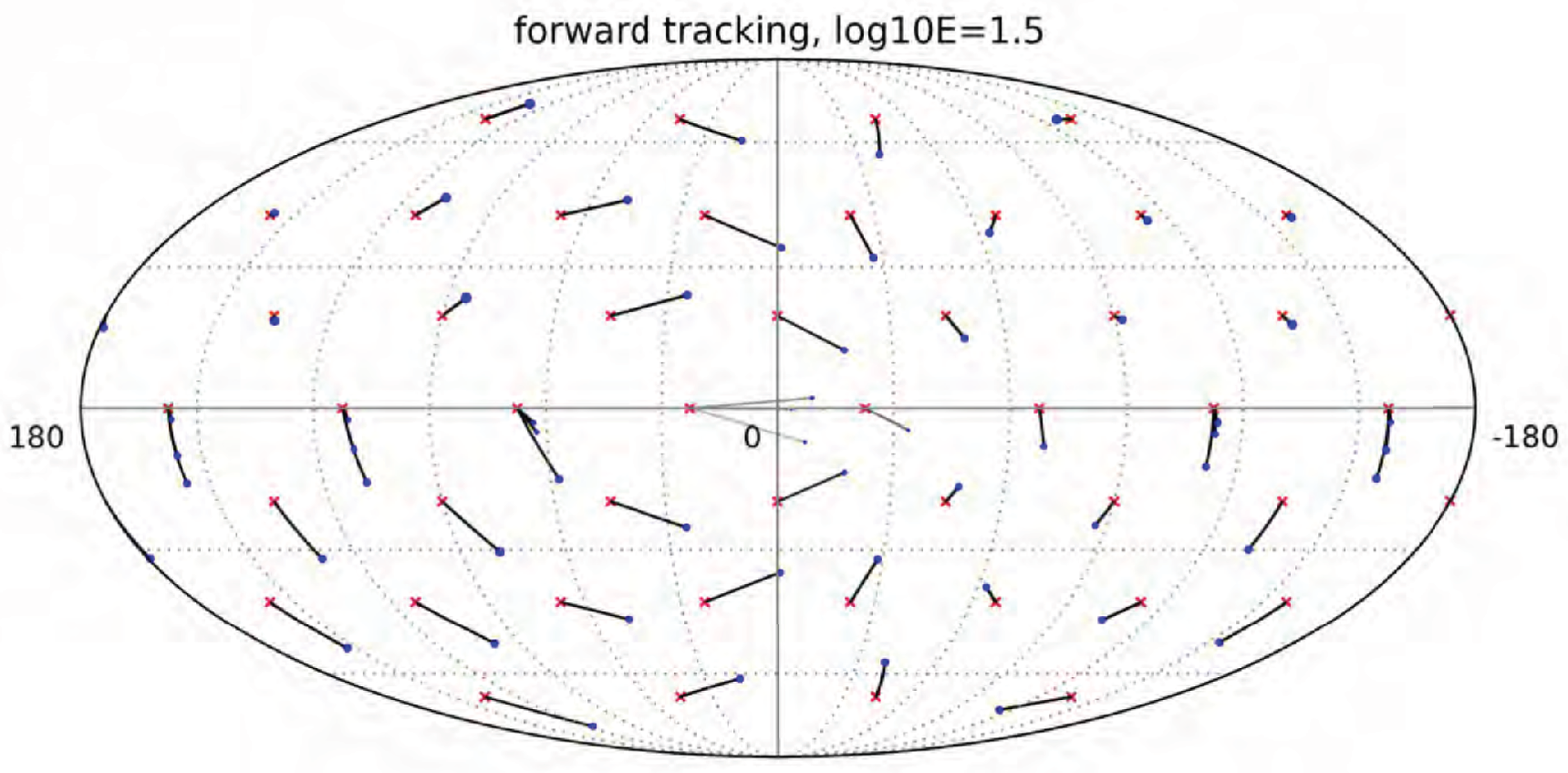}
\end{minipage}
\begin{minipage}[b]{0.48 \textwidth}
\includegraphics[width=1.15 \textwidth]{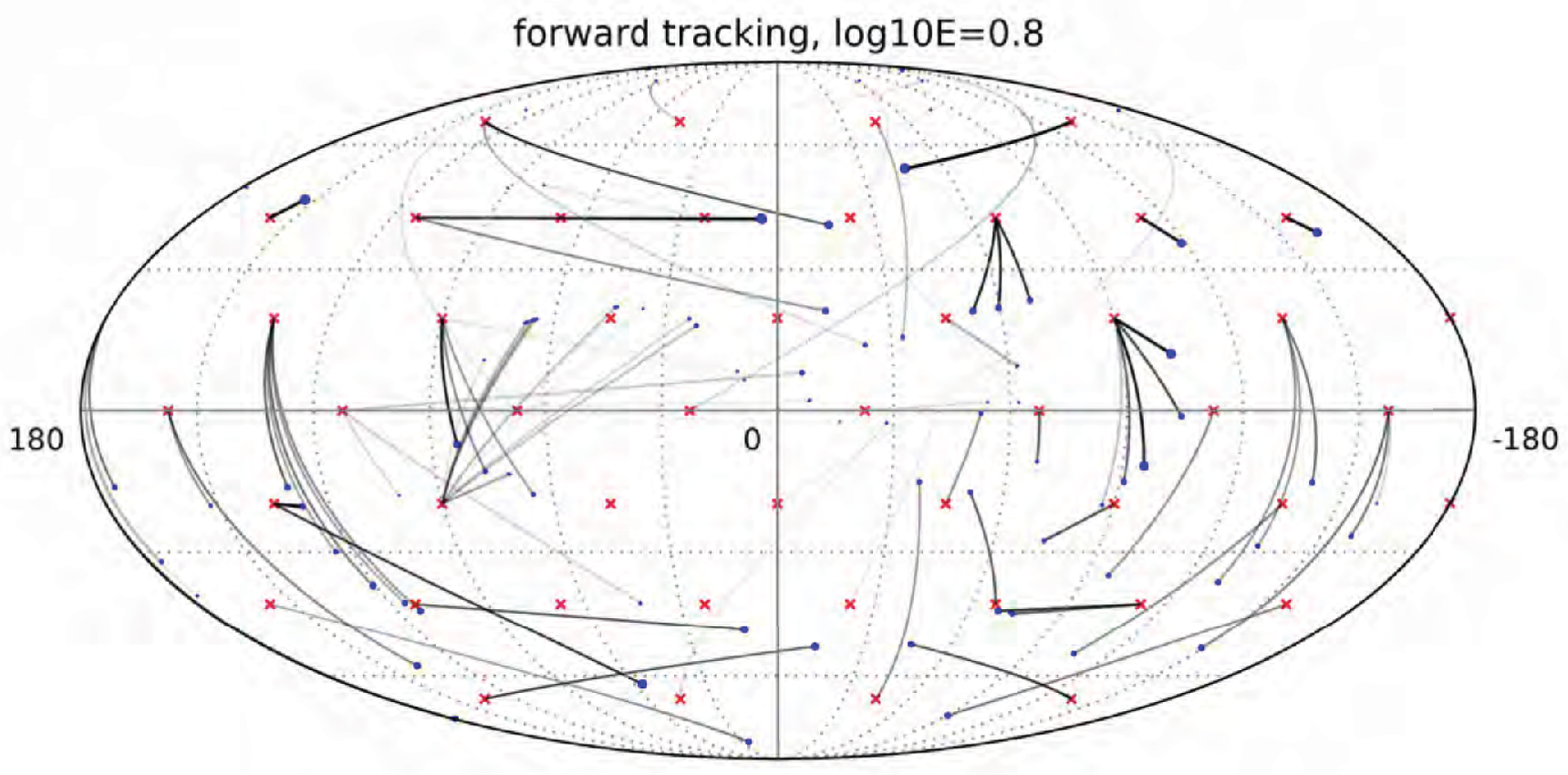}
\end{minipage}
\vspace{-0.3in}
\caption{Deflections from representative source directions (red crosses) for 32 EV (left) and 6 EV (right) CRs.  The weight of the lines reflects the relative number of CRs in the corresponding image; some sources do not illuminate Earth at the lower rigidity.} 
\label{defsreg}
\vspace{-0.1in}
\end{figure} 

Since UHECR energy losses in the Galaxy are negligible, Liouville's theorem guarantees that if the illumination from external sources is isotropic, the observed UHECR sky must be isotropic, separately for each rigidity.  Sometimes it is incorrectly argued from this, that there can be no magnification of sources, multiple imaging, etc., contrary to the expectation from lensing.  The resolution of the seeming paradox is simply that while the observed sky looks isotropic if the galaxy is illuminated isotropically, that does not mean that the observed sky is illuminated by every source direction.  Sources in some directions can make no contribution whatsoever, or have a greatly reduced flux relative to their intrinsic emission.  The magnification factor is the flux from a standard source in the given direction, summed over its images.   Fig. \ref{magmaps} shows the magnification factor for 3.2 EV UHECRs as a function of source direction, for the JF12 coherent field only (left) and for the total of coherent field plus a realization of the random field (right).  It is striking that at lower rigidities, we become blind to sources in a large portion of the extragalactic universe.  It is also noteworthy that at these low rigidities, the random field has a qualitatively important impact.  The same qualitative features were seen in \cite{giacinti+HiZturbGMF11}, which explored the importance of random fields for deflections of heavy nuclei.

\begin{figure}
\vspace{-3.3in}
\centering
\includegraphics[width=\textwidth]{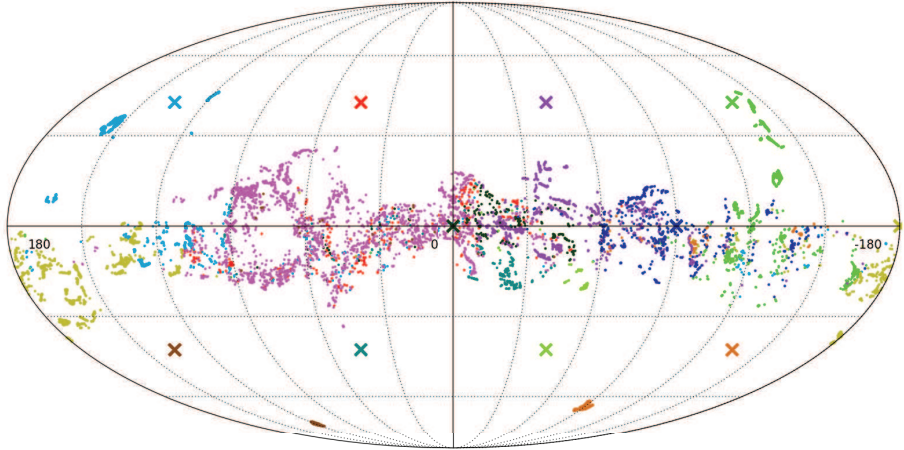}
\vspace{-3.5in}
\caption{Forward-tracked 10 EV CRs in the coherent+random field used in Fig. \ref{magmaps}, for 12 representative source directions (crosses); events from a given source have the same color as the source.  More figures showing the difference in CR deflections in different random field realizations are presented in \cite{f+CRdefs14}.  
}\label{deflogr1KRF1}
\end{figure}

The conservation of CR intensity (flux per steradian) between arrival plane and detector, enables one to replace the very computationally expensive forward-tracking from a given source direction, with high-resolution, all-sky backtracking, to make high-resolution all sky deflection maps\footnote{Using this technique, all-sky forward tracking in the JF12 regular-field at 0.1\dg\ resolution, for log$_{10}$R$_{EV}$ from 0.0 to 2.0 in steps of size 0.01-0.05, and also with several different realizations of the random field for log$_{10}$R$_{EV}$ = 0.5, 1.0 and 1.5, has been performed.  An online tool employing fast database techniques is available at http://cosmo.nyu.edu/Astroparticle, so a user can select sources and rigidities for forward-tracking.  As the backtracking for new rigidities and fields are completed, the database will be updated.   For further details see \cite{f+CRdefs14}.}.  Fig. \ref{defsreg} shows the complementary information to Fig. \ref{magdefscompare}, namely the actual deflections in the JF12 regular field, for representative source directions and two illustrative rigidities.   The weight of each line is proportional to the flux in each image, given a reference flux from the source.  The demagnification of sources is reflected here by sources with no images.  
Fig.  \ref{deflogr1KRF1} shows the observed arrival directions (for 12 selected source directions, for clarity) at 10 EV, when a realization of the random field is added to the coherent field.  One sees that the Galactic plane region is sort of an attractor for most of the 12 source directions, although for the two in the south and away from the galactic center, the Galactic plane ``repels'' the images and the total observed flux is reduced.  Further discussion will be found in \cite{f+CRdefs14}.  

\section{Conclusions and Outlook}

The present generation of GMF models has laid a good foundation for the next generation of modeling efforts, and we can expect significant improvements on a few-year timescale.  The most urgent need is to improve the models of the spatial distribution of cosmic ray electrons in the 10 GeV energy range, \ncre, and of the thermal electron distribution, \neth, which play a crucial role in determination of the Galactic magnetic field.  In parallel, advances in our theoretical insight should allow better functional forms of the GMF to be developed.  

Principal features of the most comprehensive modeling effort to date,  the JF12 model of Jansson and Farrar \cite{jf12a,jf12b} are: \\
1) A northerly-directed poloidal component whose strength is $\approx 5 \mu$G at the Galactic center and which drops to $\approx 0.2 \mu$G at the solar radius; it falls off slowly with distance from the disk.\\
2) Oppositely-directed toroidal fields in the halo, whose strength peaks at about 1 $\mu$G, $\approx 1$ kpc above and below the disk at the solar radius.\\
3) Coherent fields in a disk of $\approx$ 300 pc thickness with field strengths of up to 3 $\mu$G, in a (very grossly) 2-arm structure reversing sign between the arms and displaying variations of strength within them. (This general form is not new.) \\   
4) Random fields in both halo and disk, with significant variation from arm to arm in the disk.  The detailed structure and maximum field strength cannot be reliably determined until a better model of \ncre\ is available, but a best-estimate for the maximum strength is of order 10 $\mu$G.\\
5) Small scale anisotropic structure.  This may be due to compression and rarefaction of the coherent field, e.g., supernovae shocks, or due to stretching or compression of a random field, e.g., by winds or shocks.   Phenomenologically, both mechanisms enhance polarized synchrotron emission and are not distinguishable within the JF12 analysis; they are treated as a separate component called a striated field\cite{jf12a}.  

The qualitative features of JF12 allow several important conclusions to be drawn regarding UHECR deflections.  Most importantly, at rigidities $\approx 3-10 $ EV and below, many potential source directions are probably invisible.  Even for rigidities of order 10 EV (as for a 60 EeV carbon nucleus), deflections are large and helter-skelter for many source directions, as shown in Figs. \ref{defsreg} and \ref{deflogr1KRF1}.   Only CRs whose deflection can be reliably predicted should be used for correlation studies.  Thus being able to estimate the charge of individual UHECRs is virtually essential for source identification studies, in order for a rigidity assignment to be made\footnote{Some events will have a more reliable rigidity determination than others, and the potential accuracy on the charge assignment will depend on the composition mix.  For instance, if the composition were a simple mix of half proton and half iron, whose distributions of depth of shower maximum are very dissimilar, many events would have unambiguous charge assignments and source studies could be restricted to events whose deflections are well-determined.  Independent of how simple or complex is the composition, as long as there is a protonic component some events will have such a deep $X_{\rm max}$ that they will be unambiguously identifiable, and among those the ones with small deflection uncertainties can be selected for correlation studies.  
}.   

Today, we can unambiguously determine the rigidity for only a few UHECRs, and the GMF is only partially understood.  Consequently, source identification studies must adopt some composition ansatz and only the subset of events with relatively well-determined deflections should be used.  Within a few years, however, the quality of GMF modeling and the estimation of individual UHECRs' charges should improve significantly, greatly strengthening the prospects for constraining UHECR sources, and in the much longer run, using UHECRs as an important probe of the GMF.

\section*{Acknowledgements}
I am especially indebted to Ronnie Jansson for many years of stimulating and fruitful collaboration in developing our model of the GMF.  I also acknowledge the contributions to this review of Azadeh Keivani, Jon Roberts and  Mike Sutherland for collaboration on the deflection mapping work and for allowing me to use figures from our forthcoming paper on the topic, and David Benyamin for calculating $n_{\rm cre}$ in the model of \cite{benyamin+13}.  Finally, I wish to thank them and the many other persons who have generously contributed to my understanding of the GMF, interstellar medium,  Galactic cosmic rays, RM and synchrotron data, and related issues, including Markus Beck, Rainer Beck, Peter Biermann, Axel Brandenburg, Jim Cordes, Paolo Desiati, Klaus Dolag, Torsten Ensslin, Carmelo Evoli, Bryan Gaensler, Daniele Gaggero, Dario Grasso, Andrei Gruzinov, Marijke Haverkorn, Carl Heiles, Deepak Khurana, Luca Maccione, Ann Mao, Nir Shaviv, Frank Shu, Andy Strong, Tobias Winchen, 
and Ellen Zweibel.
This research has been supported by the National Science Foundation and NASA under grants NSF-PHY-1212538 and NNX10AC96G, and two NASA High End Computing grants of time on the Pleiades supercomputing cluster to carry out the high resolution, multi-rigidity backtracking studies.

\def\apj{Astrophys.\ J.}
\def\nat{Nature}
\def\apjl{Astrophys.\ J. Lett.}
\def\aap{Astron.\ Astrophys.}
\def\prd{Phys. Rev. D}
\def\physrep{Phys.\ Rep.}
\def\mnras{Month. Not. RAS }
\def\araa{Annual Rev. Astron. \& Astrophys.}
\def\aapr{Astron. \& Astrophys. Rev.}
\def\apss{Astrophys. \& Space Sci.}


\end{document}

 \begin{figure}
\centering
\begin{minipage}[b]{0.48 \textwidth}
\vspace{-0.1in}
\centering
\includegraphics[width=\textwidth]{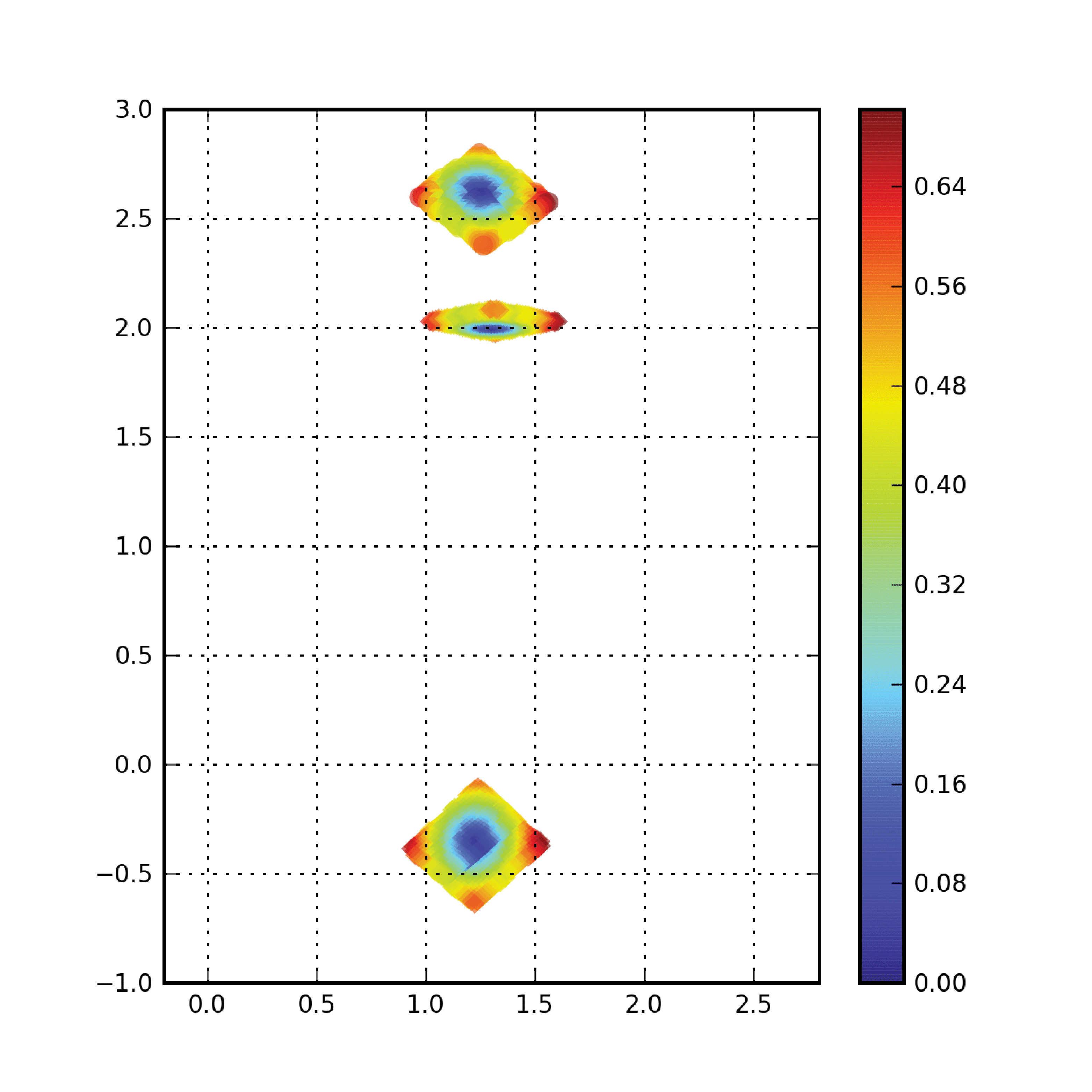}
\vspace{-0.1in}
\caption{The arrival plane regions contributing to detected 60 EV CRs for a source at \{L=341.7\dg, B=4.2\dg\}; their diamond-shapes reflect the pixelazation of the arrival direction binning.  Photons arrive at (0,0) and distanes are in kpc.   Color shows the angular distance of the UHECR from the center of its image.  }\label{arrivalplane}
\end{minipage}
\begin{minipage}[b]{0.48 \textwidth}
\vspace{-0.1in}
\centering
\includegraphics[width=\textwidth]{figs/AzadehKRF1logR1.0.eps}
\vspace{-0.1in}
\caption{Forward-tracked 10 EV CRs in the coherent+random field used in Fig. \ref{magmaps}, for illustrative source directions (crosses); events from a given source have the same color as the source.   
}\label{deflogr1KRF1}
\end{minipage}
\end{figure}

\begin{wrapfigure}{l}{0.4\textwidth}
\vspace{-1.4in}
\hspace{-0.4in}
\vspace{-1.4in}
\includegraphics[width=0.5 \textwidth]{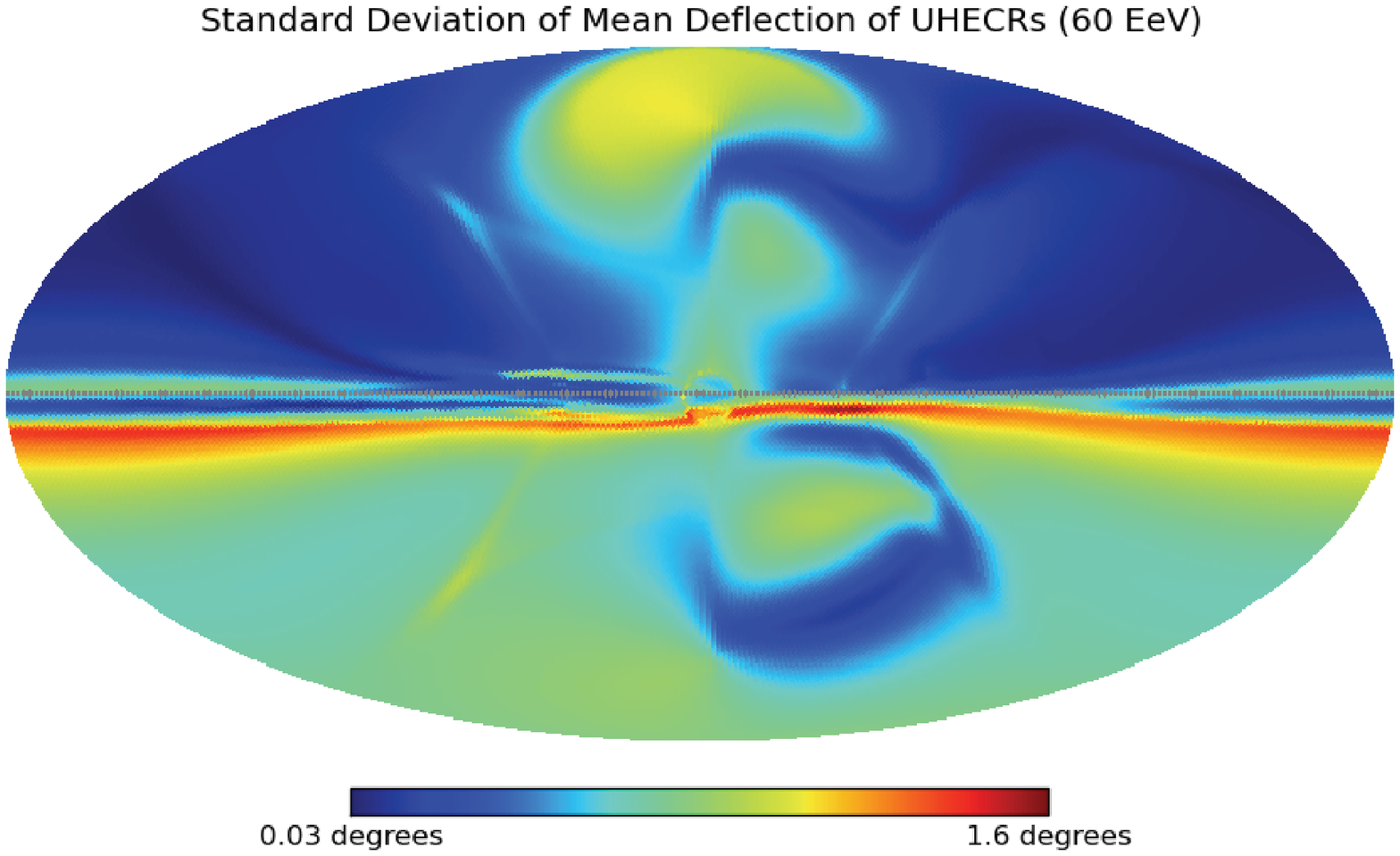}
\vspace{-0.3in}
\caption{The {\em uncertainty} in degrees on the deflections of 60 EV UHECRs shown in Fig. 1a, stemming from the parameter uncertainties in the coherent JF12 GMF, plot courtesy D. Wilmes. }\label{wilmes}

\begin{figure} 
\centering
\includegraphics[width=\textwidth]{figs/electrons.eps}
\caption{The distribution of 10 GeV electrons, ${ln(n_{\rm cre}/n_{\rm cre,\odot})}$, predicted in the source-model of ref. \cite{benyamin+13}, courtesy D. Benyamin.} 
\label{ncreBenyamin}
\end{figure} 

\end{wrapfigure}

\begin{wrapfigure}{l}{0.5\textwidth}
\centering
\includegraphics[width=0.48\linewidth]{figs/JF12disk_rand&reg.eps}
\caption{\emph{Left panel:} The random field in the disk.   \emph{Center panel:} The disk component of the  JF12 coherent field model for comparison;  it is clockwise in rings 3-6 and counterclockwise in 1,2,7, and 8. }\label{B_disk}
\end{wrapfigure}

\begin{wrapfigure}{l}{0.38\textwidth}
\centering
\includegraphics[width=0.36 \textwidth]{figs/electrons.eps}
\caption{The distribution of 10 GeV electrons, ${ln(n_{\rm cre}/n_{\rm cre,\odot})}$, predicted using the source-model of ref. \cite{benyamin+13}, from D. Benyamin.} \label{ncreHUJI}
\end{wrapfigure} 

\begin{figure} 
\centering
\begin{subfigure}[b]{0.6 \textwidth}
\includegraphics[width=\textwidth]{figs/JF12disk_rand&reg.eps}
\caption{JF12 disk field: a) random and b) regular: clockwise in rings 3-6 and counterclockwise in 1,2,7, and 8. } 
\label{B_disk}
~
\begin{subfigure}[b]{0.3 \textwidth}
\includegraphics[width=\textwidth]{figs/electrons.eps}
\caption{The distribution of 10 GeV electrons, ${ln(n_{\rm cre}/n_{\rm cre,\odot})}$, predicted using the source-model of ref. \cite{benyamin+13}, from D. Benyamin.} 
\label{ncreHUJI}
\end{subfigure} 
\end{figure}